\documentclass[aps,pra,twocolumn,superscriptaddress,nofootinbib,amsmath, amssymb]{revtex4-2}

\usepackage{graphicx}
\usepackage{dsfont}
\usepackage{slashed}

\usepackage{xcolor}

\begin{document}
	
	
	\title{Weather Forecast for Vacuum Fluctuations in QED}
	
	

\author{Maximilian Koegler}
\email{m.koegler@physik.uni-muenchen.de}
\affiliation{Arnold Sommerfeld Center for Theoretical Physics, Theresienstra{\ss}e 37, 80333 M\"unchen\\}

\author{Marc Schneider}
 \email{marc.schneider@ehu.eus}
 \affiliation{UPV/EHU University of the Basque Country, Barrio Sarriena s/n, 48940 Leioa, Spain}

\date{\today}

\begin{abstract}
We provide closed analytic expressions for the Uehling and Serber contributions of the vacuum fluctuations in QED using Meijer G-functions. Our work extends recently found results by offering a novel formulation for the Uehling and Serber potentials and explores their properties in more detail. The form of these potentials is analyzed, and their relevance for precision measurements in experiments is investigated. For the Uehling potential, we connect the solution with the propagation of photons through atmospheric turbulence. 

\end{abstract}

\maketitle

\section{Introduction} 
The hydrogen atom is one of the most intensively and accurately studied systems in theoretical and experimental physics \cite{pach96,ha05}. In quantum mechanics, the hydrogen atom's kinematics is ruled by the Coulomb potential $V_C(r)=-\frac{\alpha}{4\pi r}$ which determines the bound-state structure and the related energy levels of the electron. The vacuum state in quantum electrodynamics (QED) differs from the purely quantum mechanical description of the vacuum. The consequence are several long-range modifications of the Coulomb potential that can be triggered e.g. by strong electric fields applied to the atomic bound state \cite{haram2020transverse,lozano2020electron,petrovic2023ionization}. In these setups, the external field lowers the Coulomb barrier such that the bound-state electron can reach higher quantum numbers up to ionization, showcasing an interesting plasticity of the atomic potential\footnote{Besides this, there exists a variety of QED effects that can be probed in different setups than the hydrogen atom, e.g. in Bose-Einstein condensates \cite{brennecke2007cavity}, or exotic atoms \cite{dubrovskaya2019quantum,khetselius2018relativistic}.}.

One of the first and most remarkable hints that the Coulomb potential suffers from QED corrections was discovered because the predicted values from quantum mechanics did not match the measured energy levels. This feature of the spectrum, known as the Lamb shift \cite{lamb47} could only be understood by the QED effects: electron self-energy, anomalous magnetic moment, and vacuum polarization \cite{pauli,weis34,bethe1950numerical,pain2018vacuum,frolov2021uehling,flynn2025vacuum,roesel1977vacuum}. The last contribution describes that the bound-state electron experiences a slightly different charge than the nominal charge of the proton. The physics of QED predict that the vacuum features virtual particles that interact with the photons that form the Coulomb potential leading to a slight weakening.

Through a description as a potential, Uehling \cite{uehling} was able to derive the shift in energy levels from the vacuum polarization. At the same time, Serber \cite{serber} generalized this approach to time-dependent fluctuations. Albeit being major discoveries, one downside of their developments was the implicit integral representation of these potentials which complicates exact analyses and cloaks physical intuition.

To explore the foundations of the QED vacuum from the perspective of atomic physics, complementary to numerical methods as in \cite{dyall1989grasp, Shabaev:2015vmw}, an explicit functional form seems inevitable. In this article, we provide a novel formulation of the Uehling and Serber potentials using Meijer G-functions. It should be noted that Meijer G-functions are widespread, for instance, for various physical setups cf. \cite{antipin2019resummation,pishkoo2013some}, or \cite{kabe1958some} for the role of the Meijer G-function in statistics.

While the Uehling potential has been recently derived in analytic form \cite{medeiros2018effects}, our work provides a new perspective by extending this solution to the Serber potential and exploring their combined implications. In particular, we solve the integrals that are derived in the original reference \cite{uehling} and, thus, provide a compact, single term representation of the Uehling potential. Furthermore, we confirm that these solutions align with experimental data and agree with common approximations. The explicit form provides a novel perspective, because the Meijer G-function that describes how QED vacuum fluctuations affect the electron, occurs in the model that describes the propagation through turbulent environments such as the atmosphere \cite{chehimi2025reconfigurable, andrews2007pdf, sandalidis2010performance, Naik:22,jurado2011general,vellakudiyan2015performance} or water \cite{kumar2024meijer}. We show explicitly, that the propagation of photons within the QED vacuum and their probability to scatter at atmospheric density fluctuations are both ruled by the same Meijer G-function, thus, painting a tentative picture that motivates to study the QED vacuum like a turbulent dielectric medium. 

In contrast, the time-dependent Serber potential can be proven to vanish rapidly when moving away from the source \cite{zahn}. In fact, its contribution can only affect very short ranges in space and time which is in agreement with any measurements. 

Throughout the article, we use the following standard definitions: the mass of the electron will be denoted by $m_e$ and its elementary charge by $e$. We use the definition for the Sommerfeld fine-structure constant $\alpha=\frac{e^2}{4\pi\varepsilon_\textsc{0}\hbar c}$ and the Bohr radius $a_0=\frac{\hbar}{m_ec\alpha}$. If not stated otherwise, we employ spherical coordinates $(t,\vec x)$ where we denote the spatial part exclusively by $\vec x=(r,\vartheta,\varphi)$ and we work in the mostly negative signature for the metric.
All quantities in the main part are expressed in units of the Compton wavelength $\lambda=\frac{\hbar}{m_e c}$, such that e.g. $r\to\frac{r}{\lambda}$ which is consistent with the unit system of Serber \cite{serber}, where additionally $\hbar=c=\varepsilon_0=1$.

\section{Vacuum Effects}\label{secVacuum}
At first principles, vacuum fluctuations are expressed as a renormalized current density $\langle j^a(x)\rangle_{\rm ren}$, obtained using standard renormalization techniques, e.g. dimensional \cite{bollini1972dimensional} or Pauli-Villars regularization \cite{pauli-villars} or algebraic quantum field theoretic methods like the Hadamard renormalization technique for external potentials \cite{hol01,schlemmerzahn}. An elegant route to find $\langle j^a(x)\rangle_{\rm ren}$ has been developed by Schwinger in terms of the (slowly varying) external current $J^a(x)$ through \cite{schwinger1948quantum,schwinger1949quantum,schwinger1949quantum3,dyson1949radiation}
\begin{equation}\label{SF}
\langle j^a(x)\rangle_{\rm ren}=\sum_{n=1}^\infty \frac{a_n}{m_e^{2n}}\Box^nJ^a(x) \,,
\end{equation}
where $\Box=\partial_a\partial^a$ and $a_n$ are coefficients that depend on the specifics of the system and are essentially constructed from the Green's function (cf. \cite{Schwinger:1964zzb,schwinger1949quantum} for the detailed treatment). The physical idea behind \eqref{SF} is that the external current $J^a(x)$ (and variations thereof) induce a charge and current density $\langle j^a(x)\rangle_{\rm ren}$ in vacuum.  In QED, we would have $J^a(x)=\Box A^a-\partial^a\partial_b A^b$ \cite{dyson1949radiation}, in which the specifics of $A_a$ contain the system's details. How vacuum currents impact the energy levels of the atom requires to transform the contributions \eqref{SF}, in particular the $a_n$, into a potential form. 

\subsection{Uehling potential}
To study the modifications of the hydrogen atom inflicted by QED, we choose the static configuration for the potential, i.e. $\partial_tA_0=0$ and $\vec A=0$, and in particular for the hydrogen atom $A_0(r)=V_C(r)$ \cite{schlemmerzahn}. By working with static fields\footnote{In the original references, the static fields $\psi$ are derived as solutions of the radial Coulomb-Dirac equation with the appropriate choice for the gauge field \cite{gordon1928stoss,su85} while obeying the standard anti-commutation relations; the equation of motion is intimately tied to the conserved current $j^a=-ie\bar\psi\gamma^a\psi$.}, that is, a static external current, \eqref{SF} reduces (in linear order) to the form found by Uehling \cite{uehling}
\begin{equation}
    \langle j^{a}(\vec x)\rangle_{\rm ren}=\int \mbox{d}^3x' V_U(r)\Delta J^a(\vec{x}')\,.
\end{equation}
where the Uehling potential $V_U(r)$, with $r=|\vec x-\vec x'|$, has been defined through the integral \eqref{SF} as
\begin{align}
    V_U(r)=&-\frac{\alpha}{16\pi^4}\int_0^\frac{\pi}{2}\mbox{d}\varpi\cos^3(\varpi)\int\frac{\mbox{d}^3k}{k^2}e^{i\vec{k}\vec{x}}\nonumber\\
    &\hspace{2cm}\times\ln\left(1+\frac{k^2}{4}\cos^2(\varpi)\right).
\end{align}
Performing the $k$-integration in polar coordinates, the Uehling potential becomes
\begin{align}\label{uehlint}
    V_U(r)=\frac{\alpha}{4\pi^2 r}\int_0^\frac{\pi}{2}\mbox{d}\varpi\cos^3(\varpi) \mbox{Ei}\big(-2r\sec(\varpi)\big)\, ,
\end{align}
with exponential integral Ei$(x)=\int_{-\infty}^x\frac{{\rm d}s}{s}e^s$. 
The Uehling potential has been expressed analytically in \cite{medeiros2018effects} using Meijer G-functions, though with different formulations. In this work, we present two additional equivalent expressions and provide additional insights by extending the approach to the Serber potential. Therefore, we solve the integral in \eqref{uehlint} and find the explicit form for the Uehling potential to be  
\begin{equation}\label{Uehling}
    V_U(r)=-\frac{\alpha}{16\pi^2r}\mbox{G}^{4,0}_{2,4}\left(\!\left.\begin{matrix} 1,\frac{5}{2} \\ 0,0,\frac12,2\end{matrix}\right|r^2\!\right),
\end{equation}
where G${}^{4,0}_{2,4}$ denotes a Meijer G-function (cf. appendix~B for details)\footnote{Unluckily, the Meijer G-function was discovered exactly in the year after Uehling's and Serber's article \cite{meijer1936whittakersche}.}. Note, there exists another, more commonly used, integral representation which we also solve explicitly with Meijer G-functions (cf. appendix~C), however, as we will see \eqref{Uehling} allows us to connect our result straightforwardly  with other physical systems.

Now we are in the very comfortable position to study the closed form of the Uehling potential\footnote{There exists also an analytic formula in terms of recursive integrals \cite{pauli1936remarks, frolov}.} and compare it with experimental and theoretical predictions. 
\begin{figure}
    \centering
\includegraphics[width=1\linewidth]{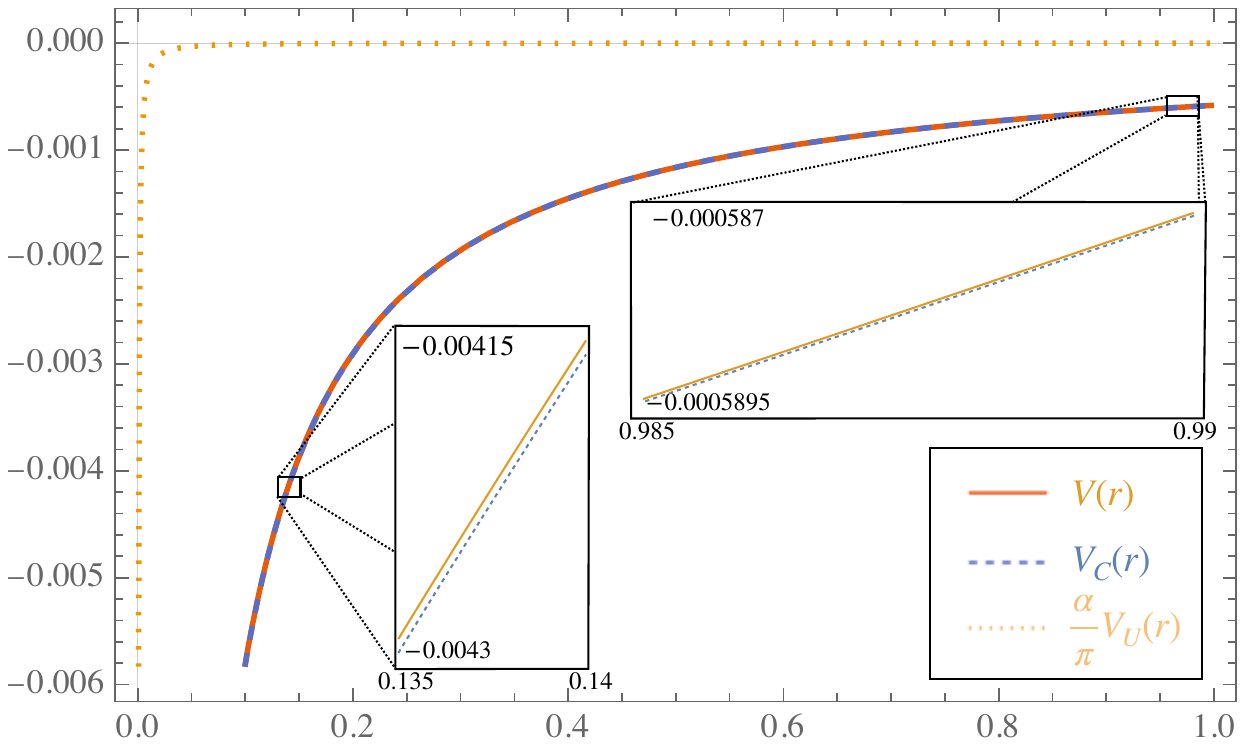}
\caption{We show the Uehling potential $\frac{\alpha}{\pi}V_U(r)$ (dotted line) against the Coulomb potential $V_C(r)$ (dashed line) in the range $r\in(0,1)$ in units of the Compton wavelength. The contribution from QED remains small compared to the external Coulomb potential. The thick orange curve $V(r)$ overlaps almost with $V_C(r)$. The zooms reveal that for a $\Delta r=0.05$ the difference between $V(r)$ and $V_C(r)$ shrinks rapidly for increasing $r$: $\frac{V}{V_C}(0.135)=0.99863$ while $\frac{V}{V_C}(0.99)=0.99994$.\label{figCU}}%
\end{figure}
Hence, we analyze the asymptotic expansions close to the origin and in the far-field limit. For small radii, i.e. $r\ll 1$,
\begin{equation}\label{Uehlingnearfield}
    V_U^0(r)\sim \alpha\left(\frac{6\big(\!\ln(r)+\gamma\big)+5}{36\pi^2 r}-\frac{1}{8\pi}+\mathcal{O}(r)\right).
\end{equation}
For $r\gg 1$, or $r\to\infty$, the Meijer G-function behaves as
\begin{equation}\label{Uehlingfarfield}
   V_U^\infty(r)\sim -\frac{\alpha}{4\pi^2}\frac{e^{-2r}}{r^{\frac52}}\,.
\end{equation}
Both asymptotic limits are in excellent agreement with the literature, and Eq.~\eqref{Uehling} additionally supplies a compact analytic expression that is valid across the full intermediate range.
Note, Uehling's derivation treats the modification separately such that the bound-state electron actually experiences the potential $V(r)=V_C(r)- \frac{\alpha}{\pi}V_U(r)$:
\begin{equation}
    V(r)=-\frac{\alpha}{4\pi r}\left[1-\frac{\alpha}{4 \pi^2}\mbox{G}^{4,0}_{2,4}\left(\!\left.\begin{matrix} 1,\frac{5}{2} \\ 0,0,\frac12,2\end{matrix}\right|r^2\!\right)\right]. \label{coulUehl}
\end{equation}  
Fig.~\ref{figCU} illustrates how the analytic Uehling potential $\frac{\alpha}{\pi}V_U(r)$ (dotted line) modulates the Coulomb potential $V_C(r)$ (straight line). The Uehling potential, as defined in \cite{uehling}, is negative but gets subtracted from the Coulomb potential, thus, shifting the potential $V_C(r)$ upwards. On the microscopic level, the photons that form the Coulomb potential, interact with the fermionic loops such that the central charge of the proton is slightly screened. This effectively reduces the charge experienced by the bound-state electron and the actual energy level is modified. Although $V_C(r)$ and $V(r)$ almost overlap for radii larger than the atomic radius, the discrepancy increases, if we approach the origin due to the logarithmic behavior of $V_U(r)$ around zero. It is understood that the probability to create fermionic loops scales with the strength of the electromagnetic background field, here $V_C(r)$. Therefore, $V_U(r)$ assumes its maximal (singular) value in the center and is otherwise a completely monotonous function on the half-line. Albeit mathematically valid, in the limit $r\to0$, the Uehling potential has to be modified once the proton of the atom is reached (in our units the proton radius $r_p=3.5\cdot10^{-4}$) \cite{ginges2016qed}. 

Putting the analytic solution to test, we derive the values of the vacuum polarization for some hydrogen energy levels and show that they agree with established theoretical predictions that are consistent with experimental data. Generically, the Uehling potential contributes to the Lamb shift\footnote{Albeit modifying the potential and the energy levels, the structural integrity of quantum mechanics remains unchallenged by QED in the sense that the postulates of quantum mechanics still hold but now with a refined Hamilton operator. This is due to the non-relativistic nature of the electron bound state.} of the hydrogen atom as follows \cite{uehling},
\begin{align}\label{Energieniveau}
	\Delta E_{nl} = \frac{\alpha}{\pi}\langle nl0|V_U|nl0\rangle =\frac{\alpha}{\pi}\!\int\! \mathrm d^3 x \, V_U(r) \left| \psi_{n l 0}(\vec x) \right|^2,
\end{align}
where the electron states $\psi_{nl0}(x)$ and a brief summary of the hydrogen atom are discussed in appendix A.\linebreak As such the values for the energy shifts using \eqref{Uehling} yield:\linebreak $\Delta E_{10} = -8.8959\cdot 10^{-7}\,\text{eV}$, $\Delta E_{20} = -1.1120\cdot 10^{-7}\,\text{eV}$, and $\Delta E_{30} = -3.2947\cdot 10^{-8}\,\text{eV}$.

These values concur with the commonly employed formula in the low momentum transfer expansion, $\Delta E_{nl}^\text{approx} = - \frac{4 m_e \alpha^5}{15 \pi n^3} \delta_{l 0}$, to within $99.69\%$ accuracy, while more refined theoretical models, such as the one in \cite{Weitz:1995zz} achieve an agreement of $99.97\%$. This approximation only shifts states with $l=0$, as detailed in \cite{greiner2008quantum}, while we report $\Delta E_{21} = -3.1658\cdot 10^{-13}\,\text{eV}$. However, this shift is two orders of magnitude smaller than the current experimental precision and thus remains beyond the reach of present measurement techniques \cite{Weitz:1995zz, Schwob:1999zz, Bezginov:2019mdi}.

The Meijer G-function and thus the Uehling potential in the compact form \eqref{Uehling} features several nice properties, for example, it is closed under various operations like the reflections $x\to-x$ and $x\to\frac1x$, scalar multiplication, Laplace and Euler transformation, and the convolution~\cite{Beals2013MeijerGA}. The Fourier transformed Uehling potential can then also be expressed in a closed form:
\begin{equation}
\begin{aligned}
	\hat V_U(k)=\frac{\alpha}{{9 \pi k^2}}& \Bigg(5-\frac{12}{k^2}+\frac{6 \left(8 - 2 k^2 - k^4\right)
   \text{arcsch}\left(\frac{2}{k}\right)}{k^3\sqrt{k^2+4}}\Bigg).
\end{aligned}
\end{equation}
This expression differs from the one presented in \cite{Frolov:2021kso} and is, in contrast, finite for $k \rightarrow 0$ and negative for all $k$. With the Fourier-transformed Uehling potential, scattering amplitudes can be conveniently determined using the Born approximation, as shown in \cite{Frolov:2021kso} for the example of Mott scattering.

Using the analytic solution of the Uehling potential, we can explore other physical systems that are described by a similar Meijer G-function to extract intuition. One example is the propagation of signals through turbulent media described by scattering processes. In this hydrodynamical picture, swirlings of the fluid alter or modulate the originally sent signal. Depending on e.g. their size they cause effects that are modeled by a Meijer G-function due to this function's generality and property to read in all necessary parameters to characterize the system.
An intriguing application thereof provides the transmission of a photon through the atmosphere to establish free-space optical quantum channels \cite{chehimi2025reconfigurable,andrews2007pdf,kumar2024meijer,jurado2011general}. The success probability of sending a photon a distance $d$ through the atmosphere is given by \cite{chehimi2025reconfigurable},
\begin{equation}
    P(d) = 1 - f(d) \, \mbox{G}^{3,1}_{2,4}\left(\!\left.\begin{matrix} 1, 1+ \vartheta(d) \\ \beta(d),\vartheta(d),\alpha(d),0\end{matrix}\right|g(d)\!\right), \label{atmo}
\end{equation}
with $f(d) = \frac{\vartheta(d)}{\Gamma(\alpha(d))\Gamma(\beta(d))}$, $g(d) = \gamma(d) \alpha (d) \beta(d)$, and the Gamma function $\Gamma(x)$. This probability incorporates atmospheric loss and pointing errors in $\gamma$ and $\vartheta$ as well as the turbulence parameters $\alpha$ and $\beta$, which represent small- and large-scale scattering cells within the atmosphere.

To draw direct comparisons with our atomic case, we calculate the intensity of the electric field originating from the potential \eqref{coulUehl}, via $E_r(r) = -\partial_r V(r)$. Utilizing identities for the Meijer G-function and its derivative from appendix A, the intensity results in
\begin{equation}
    I(r) \propto |E_r(r)|^2 \approx \frac{\alpha^2}{16\pi^4r^4}\left[1+\frac{\alpha}{\pi}\mbox{G}^{3,1}_{2,4}\left(\!\left.\begin{matrix} 1,\frac{5}{2} \\ 0,\frac32,2,0\end{matrix}\right|r^2\!\right)\right]. \label{atomIntens}
\end{equation}
Since this particular Meijer G-function is negative for all $r$ the intensity $I$ is damped due to vacuum fluctuations.

In classical electrodynamics, intensity represents the energy flux carried by the electromagnetic field, proportional to the square of the field amplitude. In quantum theory, this concept is reinterpreted: the intensity of the field becomes proportional to the probability density for detecting a photon, ($I \sim P$), as for example in the context of the double-slit experiment \cite{dirac1927quantum}. There, the interference pattern arises from the probability amplitudes associated with all possible paths, whose squared modulus gives the measurable intensity. This formal bridge between classical field intensity and quantum detection probability implies that classical damping effects—such as those modeled by the Meijer G-function in turbulent or fluctuating media—can have direct analogs in the suppression of quantum amplitudes. Consequently, the observed similarity between the expressions governing atmospheric photon propagation and the electric field of a nucleus modified by QED corrections points toward a deeper connection. Both phenomena are mediated by fluctuations—classical in the atmosphere, quantum in the vacuum—and are encoded by the same function in \eqref{atmo} and \eqref{atomIntens}, suggesting that this particular Meijer G-function captures a universal response of wave propagation through stochastic media, regardless of their classical or quantum origin.

This structural similarity suggests that the QED vacuum influences the electric field of an atomic nucleus in a manner akin to how atmospheric turbulence affects optical signal propagation. It should be mentioned that both effects are distinct in their theoretical description, while the QED vacuum fluctuations represent quantum processes, the scattering cells in the atmosphere obey classical fluid dynamics. Comparing the two expressions, \eqref{atmo} and \eqref{atomIntens}, reveals that, in the atomic context, only the small-cell turbulence parameter $\alpha$ contributes. This aligns with the known behavior of vacuum fluctuations in QED, i.e. virtual electron-positron loops contribute on scales much smaller than the photon's wavelength due to the uncertainty principle. Along this line, the analogy motivates a reinterpretation of the QED vacuum as a statistically structured medium, whose impact on field propagation may be captured through effective models reminiscent to those used in classical wave physics.

\subsection{Serber potential}
Building on Uehling's analysis, Serber generalized the approach to time-dependent external currents \cite{e2018microscopic,zahn,serber}. In this work, we derive explicit forms for both the Uehling and Serber potential using Meijer G-functions. Hence, we provide novel forms of the Uehling potential and to derive the explicit form for the Serber potential. Taking time-dependent external currents into account, the vacuum current density \eqref{SF} has to be considered in the full, four-dimensional case. Following \cite{serber}, we transform into hyperbolic coordinates in $k$-space such that, $k_0=K\sinh(\theta)$ and $k=K\cosh(\theta)$ with $\theta\in[0,\infty)$, thus, 
\begin{equation}
\begin{aligned}
	\langle j^{a}(x)\rangle_{\rm ren}=& \int \mbox{d}^4x' \int_0^{\frac\pi2}\mbox{d}\varpi \cos^3(\varpi)\\
	&\times \Lambda(x-x';\varpi)\Box\Box J^a(x') \,,\label{curdens}
\end{aligned}
\end{equation}
where $\Lambda(x-x';\varpi)$ denotes the integration kernel. Under appropriate boundary conditions, cf. \cite{serber}, the integral kernel consists of a static and a dynamic contribution, i.e.\linebreak $\Lambda(x-x';\varpi)=\Lambda_{\rm stat}(r;\varpi)+\Lambda_{\rm dyn}(t,r;\varpi)$ where
\begin{align}
\Lambda_{\rm stat}(r;\varpi)=&-\frac{\alpha}{16\pi^2r}\cos^2(\varpi)\int_1^\infty \frac{\mbox{d}K}{K^3}e^{-2Kr\sec(\varpi)}\,, \label{serberL1}\\
\Lambda_{\rm dyn}(\tau;\varpi)=&\frac{\alpha}{8\pi^2\tau}\cos(\varpi)\int_1^\infty \frac{\mbox{d}K}{K^2}\mbox{J}_1(2K\tau\sec(\varpi))\,.\label{lambda2}\;\;\;\;
\end{align}
Here, ${\rm J}_1(x)$ is the Bessel function of the first kind and we introduced the shorthand notation for the geodesic distance $\tau=\sqrt{|(r-r')^2 - (t-t')^2|}$. The boundary conditions further infer that $\Lambda_{\rm dyn}(\tau;\varpi)$ has solely support in the future directed light cone and vanishes outside \cite{serber,zahn}. 

Since we are particularly interested in the dynamical modification, we extract the Serber potential $V_S(t,r)$ by the usual method
\begin{equation}
\langle j^a(x)\rangle_{\rm ren}=\int \mbox{d}^4x' \, V_S(t,r)\Box\Box J^a(x')\,.
  \label{renormCurr}
\end{equation}
where, similar to the time-independent derivation, the potential is found through the $\varpi$-integral in \eqref{curdens}
\begin{equation}\label{serbint}
   V_S(t,r)= \int_0^{\frac\pi2}\mbox{d}\varpi \cos^3(\varpi)\Lambda_{\rm dyn}(t,r;\varpi)\,.
\end{equation}
The $K$-integral in \eqref{lambda2} can be evaluated directly to:
\begin{equation}
\Lambda_{\rm dyn}(y)\!=\!\frac{\alpha}{8\pi^2}\!\left[\frac12-\big(\!\ln(y)\!+\gamma\big)\!+\frac{y^2}{4}\,{}_2F_3\!\left(\!\left.\begin{matrix}1, \;1\\2,\; 2,\; 3\;\end{matrix}\right|\!-y^2\!\right)\right]
\label{lambda2equ}
\end{equation}
where we introduced $y:=\tau\sec(\varpi)$ for simplicity. 
\begin{figure}
  \centering
{\includegraphics[width=0.85\linewidth]{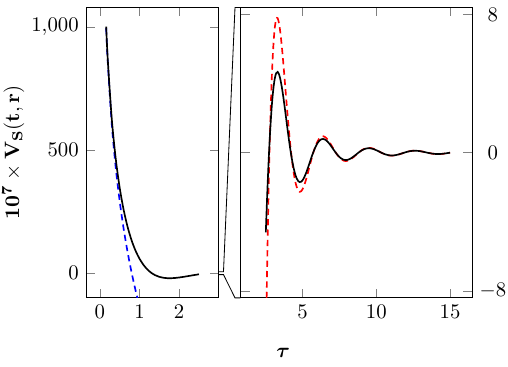}}%
\caption{Left panel: Serber potential $V_S(\tau)$ in units of the Compton wave-length for $\tau \in [ 0.16, 2.5]$ where the logarithmic divergence is dominant. The validity of the small $\tau$ approximation $V_S^0(\tau)$ is shown (blue line).
Right panel: for $\tau \in [ 2.5, 15] $, (zoom into the tail of $V_S(\tau)$) the Meijer G-function performs damped oscillations such that the Serber potential agrees with the large $\tau$ expansion $V_S^\infty(\tau)$ (red line).}\label{serbplot}%
\end{figure}
Taking \eqref{lambda2equ}, we perform the $\varpi$-integration in \eqref{serbint} and find the closed form of the Serber potential to be
\begin{equation}\label{Serber}
V_S(t,r)=\frac{\alpha}{8\pi^2} \! \! \left[\frac{\sqrt{\pi}}{4}\mbox{G}^{2,2}_{3,5}\!\left(\!\left.\begin{matrix} 1,1,\frac{5}{2} \\ 1,2,-1,0,0\end{matrix}\right|\tau^2\!\right)-\!\frac{2\ln(\tau)}{3}\!+\!c_S\right]
\end{equation}
with constant
$c_S=-\frac23(\frac13 + \gamma-\ln(2))$ and, again, we encounter a (different) Meijer G-function. To develop a first intuition about \eqref{Serber}, we plotted $V_S(\tau)$ in Fig.~\ref{serbplot}. 
Clearly, the Serber potential shows a divergence at the origin and a damped oscillation around zero towards large $\tau$ values. This can be understood from asymptotically expanding $V_S(t,r)$: for $\tau\ll1$, we can approximate 
\begin{equation}
V_S^0(t,r) \sim -{\alpha} \frac{3 \big(\!\ln(\tau) +\gamma \big) +1  - \ln(8)}{36 \pi^2}+ \mathcal{O}(\tau)\,,\label{serbnear}
\end{equation} 
finding that the divergence is logarithmic at $\tau=0$, while for $\tau\gg1$, $V_S(t,r)$ goes into a damped oscillation
\begin{equation}
    V_S^\infty(t,r) \sim  \alpha \frac{4 \tau \cos(2 \tau) + 11 \sin(2\tau)}{128 \pi^2 \tau^4} + \mathcal{O}\!\left(\frac{1}{\tau^5}\right).\label{serbfar}
\end{equation}
The asymptotic expansion for $V^0_S(\tau)$ agrees with the dynamical kernel for $\tau<0.5$ while $V^\infty_S(\tau)$ matches $V_S(\tau)$ for $\tau>8$. Note, appendix~D shows that \eqref{serberL1} vanishes asymptotically as well. 

\begin{figure}
    \centering
\includegraphics[width=0.75\linewidth]{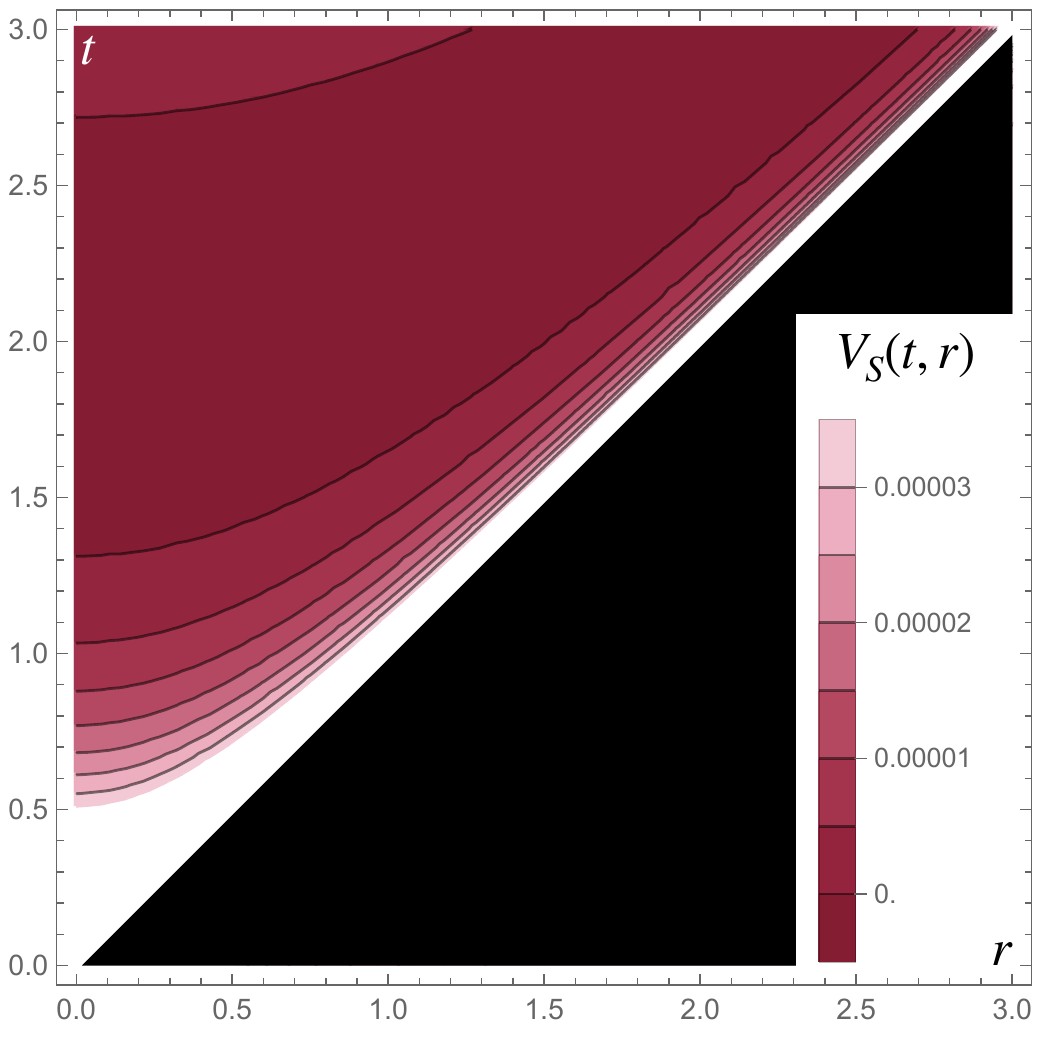}
\caption{Zoom into the region $t\in[0,3]$ and $r\in[0,3]$ for the Serber potential: The value of $V_S(t,r)$ is decoded in shades of magenta as being purely positive. Close to the light cone, the potential diverges (lighter shading) and becomes completely monotonous after the minimum along the hyperbola $\tau\approx1.75$.  
\label{figCSn}}%
\end{figure}

Nevertheless, this analysis captured only the behavior in the geodesic distance while a better understanding of $V_S(t,r)$ would be gained in terms of $t$ and $r$.\linebreak
\begin{figure}
    \centering
\includegraphics[width=0.75\linewidth]{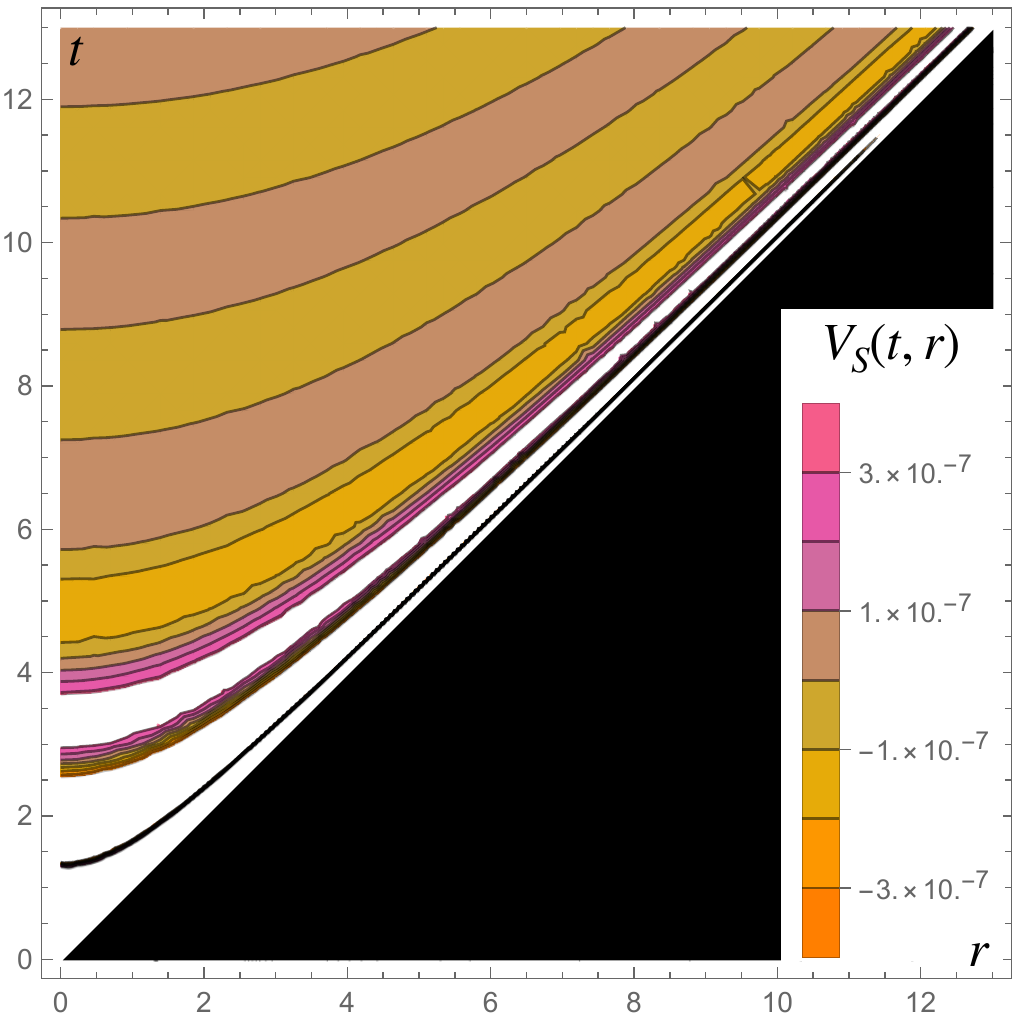}
\caption{Serber potential $V_S(t,r)$ within the future light cone, $|r-r'|^2-|t-t'|<0$, from $t\in[0,13]$ and $r\in[0,13]$: Positive values are represented by magenta hues and negative ones by orange hues. At the boundary, the potential diverges while in the far interior, we see a, oscillation of the order $2\cdot10^{-8}$. 
\label{figCS}}%
\end{figure}
Fig.~\ref{figCS} shows the support of the Serber potential within a $t$-$r$-diagram. The ceasing of the oscillations occurs very rapidly; at $t\approx20$, they reach amplitudes that challenge the numerical resolution. From Fig.~\ref{serbplot} and \ref{figCSn}, and \eqref{serbfar}, we discover that for time-spans of about two Compton times the modulation from $V_S(t,r)$ turns into damped oscillations around zero. The Serber potential describes a propagation of an electromagnetic pulse induced by the current of the fermionic loops \cite{serber}. This pulse is very localized, therefore its immediate effect can only be observed within a few Compton wave lengths which is reflected by the monotonous part of $V_S(t,r)$ while afterwards, the initial pulse dies off in a damped oscillation.

Inserting for instance $V_S(t,r)$ into \eqref{Energieniveau} as potential and choosing the ground-state wave function, we find that $\Delta E_{nl}\to0$ for $t\to\infty$ and/or $r\to\infty$. Hence, we would not expect a significant net modification of $V_C(r)$ from $V_S(t,r)$, but rather a modulation that averages out (temporally and spatially) away from the source. In fact, we are interested in the time and position, $t_\star$ and $r_\star$, for which the Serber contribution to the energy-shift
\begin{equation}
\Delta E_{10}^S(t,r)=\int\! \mathrm d^4 x \, V_S(t,r) \left| \psi_{n l 0}(\vec x) \right|^2,
\end{equation}
averages out. To estimate $t_\star$ for oscillating functions one demands that $ct_\star\gg \lambda$, where $\lambda$ is the wave-length of the oscillation \cite{cohen2024atom}. For convenience, one chooses as threshold $t_\star\ge 10\cdot \frac{\lambda}{c}$. In our case, we deduce from \eqref{serbfar}, that the frequency is related to $\lambda=\lambda_C/2\approx2.43\cdot10^{-12}$m, yielding $t_\star\approx2\cdot10^{-19}$s such that the influence of the Serber potential averages out. For the radial averaging, we consequently find $r_\star\ge 10\lambda_C\approx2.43\cdot10^{-11}$m which is consistent with the explanation in \cite{serber}. Hence, $V_S(t,r)$ contributes only on short time spans. Since QED fermions are massive, they propagate inside the light cone which confines any potential influence of $V_S(t,r)$ to the vicinity of the source where the Serber potential remains monotonous.

\section{Discussion}
Deriving the explicit forms of the Uehling and Serber potential allowed us to explore the features of the QED vacuum from a new perspective. Because hydrogen atoms are extremely accurately measured, we substantiated our solution since all predictions from the Meijer G-functions matched the results in the literature accurately. We furthermore saw that the Serber potential may only affect time-resolved measurements while averaging out effectively, whereas the Uehling contributions persists. 

In the study of the Lamb shift, as discussed by \cite{paillet2019highly},
an analytic and closed form of the Uehling potential enhances precision and accelerates numerical computations, such as in the context of superheavy elements and molecules where the Lamb shift is significantly amplified, reaching magnitudes of up to $100\,\text{eV}$ \cite{johnson1985lamb, Schwerdtfeger:2015etv, ginges2016qed, smits2023pushing}. The commonly used fitted parametric representations of the Uehling potential can now be complemented by Eq.~\eqref{Uehling}. 

The occurrence of the Meijer G-function revealed a remarkable resemblance with the propagation of photons through the atmosphere. The Uehling potential arises from the vacuum polarization correction to the photon propagator induced by virtual electron–positron pairs. These fluctuations, confined to distances on the order of the electron Compton wavelength, effectively screen the Coulomb interaction at short range, well below typical atomic scales. This can be interpreted as a scattering of the exchanged virtual photon off vacuum fluctuations, introducing a statistical modification to the interaction. A structurally analogous effect occurs in atmospheric optics, where photons scatter off turbulent inhomogeneities. In both cases—despite the quantum origin in QED and the classical origin in atmospheric turbulence—the cumulative effect of microscopic fluctuations modifies signal propagation in a way that is captured by the same Meijer G-function, encoding the statistical structure of the underlying medium. This analogy potentially offers a new perspective on how to describe the QED vacuum: not merely as a formal correction in perturbation theory, but as an effective fluctuating medium with statistical properties amenable to macroscopic modeling.

Our particular Meijer G-function, hence, supports the idea of the QED vacuum as a fluctuating environment in which tiny (by Heisenberg's uncertainty principle) inhomogeneities are created. Any propagating photon could in principle interact with these fermionic vacuum bubbles. The Uehling potential inflicts changes in the current density that sources the inhomogeneous Maxwell equations \cite{serber}. As such, one may use the closed form to facilitate the derivations of backreaction on the field strength tensor within the Maxwell equations, as suggested in \cite{fer19,pla20}.

While the cumulative net effect is captured by the Uehling potential, the Serber potential describes the influence of such a fluctuation dynamically and contributes only within a small (with respect to Heisenberg's uncertainty principle) space-time region around the event. It is understood that the QED fluctuations occur in charge rather than molecular density and therefore are subjected to two different scientific theories, one microscopic one macroscopic. To be precise, while QED fluctuations are governed by quantum field theory and occur at length scales of 1 fm up to 1 pm, atmospheric fluctuations are ruled by fluid mechanics and range from 1 cm for small cells to 1 km for large cells. However, the similarity of the solutions supports the intuition of viewing the vacuum as a sort of turbulent medium. A promising fact is that non-linear effects in QED can be traced back to such loop couplings as well \cite{karb15,karb19}. How far the analogy can be spun, has to be investigated in further research.

Although the Uehling and the Serber potential represent only the first order QED correction they are essential in the context of the hydrogen atom. However, for atoms with higher atomic number or muonic atoms the Wichmann-Kroll potential (third order correction) becomes important \cite{wichkroll} in the vacuum polarization.

The Lamb-shift, however, represents only one of various long-range modifications that affect the Coulomb potential. In fact, external electromagnetic fields, are known to deform the Coulomb potential with observable consequences for atomic and molecular physics \cite{haram2020transverse,indelicato2019qed,fedotov2023advances}. Especially when strong fields act on the atomic bound-state, the Coulomb barrier is reduced in a certain direction, thus leading to high harmonics generations and an enhanced ionization probability \cite{lozano2020electron,petrovic2023ionization}. If the external field is chosen to be dynamical, one can show that the induced motion of the nucleus leads to a sort of microscopic dynamical Schwinger effect \cite{e2018microscopic} that influences the ionization rate as well. In non-linear QED, the Uehling and the Serber potential enter the Euler-Heisenberg Lagrange density \cite{heisenberg1936folgerungen}, thus, influencing photon-photon scattering, photon splitting, and other effects. The closed analytic forms help to tackle these questions in the future.

\section*{Acknowledgments}
We want to express our gratitude to Cecilia Giavoni and Stefan Hofmann for fruitful comments and suggestions on the early version of this article. We thank the referees for their constructive and valuable comments and suggested references that significantly improved this manuscript. This work has been supported by the Basque Government Grant IT1628-22 and by the Grant PID2021-123226NB-I00 (funded by MCIN/AEI/10.13039/501100011033 and by ``ERDF A way of making Europe'').


\appendix
\section*{Appendix}

\section{Hydrogen atom.} 
The electron bound state of the hydrogen atom is given as the solution to the Schrödinger equation $ i\partial_t|\Psi\rangle_t=H|\Psi\rangle_t$. In SI-units,
\begin{equation}
    i\hbar\partial_t\Psi(t,\vec x)=H\Psi(t,\vec x)=\!\left(-\frac{\hbar^2}{2m_e}\Delta_{\mathbb{S}_3}+\frac{e^2}{4\pi \varepsilon_0r}\right)\!\Psi(t,\vec x),
\end{equation}
where we called the Hamilton operator $H$ and the spherical Laplace operator $\Delta_{\mathbb{S}_3}$. The wave-functions in position space $\Psi(t,\vec x)=\langle \vec{x}|\Psi\rangle_t=\langle\vec{x}|nlm\rangle_t=e^{-\frac{i}{\hbar}E_{n}t}\psi_{nlm}(\vec x)$ read
\begin{equation}
    \psi_{nlm}(\vec x)=R_{nl}(r)Y_{ml}(\vartheta,\varphi)
\end{equation}
with spherical harmonics $Y_{ml}(\vartheta,\varphi)$ and radial part 
\begin{equation}
  R_{nl}(r)=N_{nl}\;e^{-\frac{r}{na_\textsc{0}}}\left(\frac{2r}{na_0}\right)^lL^{2l+1}_{n-l-1}\!\left(\frac{2r}{na_0}\right).
\end{equation}
The associated Laguerre polynomials are defined according to \cite{heuser2013gewohnliche} which is consistent with those used by \texttt{mathematica 14.1} and the normalization
\begin{equation}
    N_{nl}=\sqrt{\left(\frac{2}{na_0}\right)^3\frac{(n-l-1)!}{2n(n+l)!}}\,.
\end{equation}
The energy levels are determined by the eigenstates of the Hamilton operator $H\psi_{nlm}(\vec x)=E_n\psi_{nlm}(\vec x)$ to be $E_n=-\frac{\rm Ry}{n^2}$ with Rydberg energy Ry $=\frac{e^4m_e}{32\pi^2\varepsilon_\textsc{0}^2\hbar^2}$. For convenience, we will present the most important states here, that will be used in the article: the s-states
\begin{align}
\psi_{100}(\vec x) &= \frac{1}{\sqrt{\pi a_0^3}} e^{ - \frac{r}{a_\textsc{0}}}\,,\label{100} \\
\psi_{200}(\vec x) &= \frac{1}{4\sqrt{2\pi a_0^3}} \left(2-\frac{r}{a_0}\right) e^{ - \frac{ r}{2a_\textsc{0}}}\,,\label{200}\\
\psi_{300}(\vec x) &= \frac{1}{81\sqrt{3\pi a_0^3}} \left(27-\frac{18r}{a_0}+\frac{2r^2}{a_0^2}\right) e^{ - \frac{ r}{3a_\textsc{0}}}\,,\label{300}
\end{align} 
and the first p-state
\begin{equation}
\psi_{210}(\vec x) = \frac{\cos(\vartheta)}{4\sqrt{2\pi a_0^3}} \frac{r}{a_0} e^{ - \frac{ r}{2a_\textsc{0}}}\,,\label{210}
\end{equation}
which is, without Lamb shift, energetically degenerate with the 2s-state.

\section{Meijer G-function.} 
The Meijer G-function solves the differential equation $Df=0$ where 
\begin{equation}\label{MDG}
D=\prod_{j=1}^q\left(x\frac{{\rm d}}{{\rm d}x}-b_j\right)-(-1)^{p-m-n}x\prod_{j=1}^p\left(x\frac{{\rm d}}{{\rm d}x}+1-a_j\right)
\end{equation}
with $x\in\mathbb{C}$, indices $\{a_j\}$ and $\{b_j\}$, and non-negative integers $n$, $m$, $p$, $q$. For $0\le n\le p$ and $0\le m\le q$, Meijer found the solution by an inverse Mellin transformation \cite{meijer1936whittakersche,Beals2013MeijerGA}, that is the line-integral 
\begin{align}
f(x)&= \mbox{G}^{m,n}_{p,q}\left(\!\left.\begin{matrix} a_1,\ldots,a_n,a_{n+1},\ldots,a_p \\ b_1,\ldots,b_m,b_{m+1},\ldots,b_q\end{matrix}\right|x\!\right)\\
=&\frac{1}{2\pi i}\int_\mathcal{C}\frac{\prod_{j=1}^m\Gamma(b_j-s)\prod_{j=1}^n\Gamma(1-a_j+s)x^s}{\prod_{j=m+1}^q\Gamma(1-b_j+s)\prod_{j=n+1}^p\Gamma(a_j-s)}\mbox{d}s\nonumber\label{MeijerG}
\end{align}
along one path $\mathcal{C}$ within the complex plane. The Meijer G-function features several nice properties, for example, it is closed under various operations like the reflections $x\to-x$ and $x\to\frac1x$, scalar multiplication, Laplace and Euler transformation, and the convolution \cite{Beals2013MeijerGA}. One remarkable property is that many elementary functions can be expressed as Meijer G-function, e.g. 
\begin{equation}
    e^x=\mbox{G}^{0,1}_{1,0}\!\left(\!\left.\begin{matrix}  \\ 1\end{matrix}\right|-x\!\right),\hspace{0.5em}\mbox{or}\hspace{0.5em}  \ln(x+1)=\mbox{G}^{1,2}_{2,2}\!\left(\!\left.\begin{matrix} 1,1 \\ 1,0\end{matrix}\right|x\!\right).
\end{equation}
Moreover, the Meijer G-function satisfies a plethora of interesting identities; the following have been used in this article. The first is a very elementary one for $\ell\in\mathbb{Z}$,
\begin{align}\label{meijertrafo1}
    \mbox{G}&^{m,n}_{p,q}\left(\!\left.\begin{matrix} a,a_2,\ldots,a_n,a_{n+1},\ldots,a_p \\ b_1,\ldots,b_m,b_{m+1},\ldots,b_{q-1},a+\ell\end{matrix}\right|x\!\right)\\
    =&\,(-1)^\ell\mbox{G}^{m+1,n-1}_{p,q}\left(\!\left.\begin{matrix} a_2,\ldots,a_n,a_{n+1},\ldots,a_p,a \\ a+\ell,b_1,\ldots,b_m,b_{m+1},\ldots,b_{q-1}\end{matrix}\right|x\!\right),\nonumber
\end{align}
which allows to exchange the indices while changing the parameters. Another identity involves the derivative
\begin{align}
\frac{\rm d}{{\rm d}x}&\left[x^{-b_1}\;\mbox{G}^{m,n}_{p,q}\left(\!\left.\begin{matrix} a_1,\ldots,a_n,a_{n+1},\ldots,a_p \\ b_1,\ldots,b_m,b_{m+1},\ldots,b_q\end{matrix}\right|x\!\right)\right]\\
&=-x^{-(b_1+1)}\;\mbox{G}^{m,n}_{p,q}\left(\!\left.\begin{matrix} a_1,\ldots,a_n,a_{n+1},\ldots,a_p \\ b_1+1,\ldots,b_m,b_{m+1},\ldots,b_q\end{matrix}\right|x\!\right).\nonumber
\end{align}
Others can be found in standard collections of tables, integrals, and special functions, e.g. \cite{luke1969special}. 

\section{Other representation.} Several other integral representations of $-\frac{\alpha}{\pi}V_U(r)$ exist, e.g. the one by Uehling \cite{uehling}
\begin{align}\label{VUUehl}
    \bar{V}_U(r)=-\frac{2\alpha^2}{3\pi r}\!\int_1^\infty\!\!\mbox{d}x\,e^{-2\frac{rx}{\lambda}}\frac{2x^2+1}{2x^4}\sqrt{x^2-1}\,.
\end{align}
Again, this integral is solvable in terms of Meijer G-functions
\begin{align}\label{VUaD}
 &\int_1^\infty\!\!\!\!\mbox{d}x\,e^{-2 \frac{rx}{\lambda}}\frac{2x^2+1}{2x^4}\sqrt{x^2-1}=\frac{3}{2 \pi} \frac{r}{\lambda}+\frac{2}{3 \pi} \frac{r^3}{\lambda^3}\nonumber\\
 &\hspace{0.25em} +\frac{1}{2}\mbox{G}^{2,0}_{2,4}\!\left(\!\left.\begin{matrix} \frac12,\frac{3}{2} \\ 0,0,\frac12,\frac12\end{matrix}\right|\frac{r^2}{\lambda^2}\!\right)+\frac{1}{4}\mbox{G}^{2,0}_{2,4}\!\left(\!\left.\begin{matrix} \frac{1}{2},\frac{5}{2} \\ 0,1,\frac12,\frac12\end{matrix}\right| \frac{r^2}{\lambda^2}\!\right).
\end{align}
To verify our result, we numerically confirmed that the difference $\bar V_U(r)-(-\frac{\alpha}{\pi})V_U(r)$, using \eqref{Uehling} and \eqref{VUaD}, yielded zero. From \cite{uehling}, we know that \eqref{uehlint} can be represented by \eqref{VUUehl} which in turn leads to a novel identity for the Meijer G-function and a representation of polynomials in terms of Meijer G-functions. In addition, the behavior of the Meijer G-function under Fourier transform allows to verify analytically that the Fourier transform $-\frac{\alpha}{\pi}\hat{V}_U(k)$ and $\hat{\bar{V}}_U(k)$ are equal. 

\section{Kernel asymptotics}
To understand the fading impact of the Serber contribution at long distances, let us analyze the two kernels \eqref{serberL1}, and \eqref{lambda2} directly. We consider first the dynamical kernel
\begin{equation}
    \Lambda_{\rm dyn}(y)\!=\!\!\frac{\alpha}{8\pi^2}\!\left[\frac12-(\ln(y)+\gamma)\!+\!\frac{y^2}{4}{}_2F_3\!\left(\!\left.\begin{matrix}1, \;1\\2,\; 2,\; 3\;\end{matrix}\right|-y^2\!\right)\right]
\end{equation}
where the hypergeometric function for large arguments $y\to\infty$,
\begin{equation}
    {}_2F_3\left(\!\left.\begin{matrix}1, \;1\\2,\; 2,\; 3\;\end{matrix}\right|-y^2\!\right)\sim\frac{4}{y^2}\left(\ln(y)+\gamma-\frac12+\mathcal{O}\left(\frac{1}{y^\frac52}\right)\right).
\end{equation}
Comparison with $\Lambda_{\rm dyn}(y)$ shows that these terms exactly cancel and the contribution goes asymptotically to zero.

The same conclusion holds true for the static part for which the $K$-integral yields 
\begin{align}\label{lamstat}
    \Lambda_{\rm stat}(r;\varpi)=&\frac{\alpha}{16\pi^2 r}\cos^2(\varpi)\Big[e^{-2r\sec(\varpi)}\big(1-2r\sec(\varpi)\big)\nonumber\\
    &+ \, r^2\sec^2(\varpi)\mbox{E}_1\big(r\sec(\varpi)\big)\Big],
\end{align}
involving the exponential integral E${}_1(x)=\int_1^\infty$d$s\frac{e^{-sx}}{s}$. For large radii, it is clear that the first two terms are suppressed by the damping exponential and we are left with a contribution $x$E${}_1(x)$. By using the representation through Tricomi's confluent hypergeometric function U$(a,b,x)$ one can derive the asymptotics
\begin{equation}
    \mbox{E}_1(x)=e^{-x}\;\mbox{U}(1,1;x)\overset{x\to\infty}{\longrightarrow}\frac{e^{-x}}{x}\left(1+\mathcal{O}\left(\frac{1}{x}\right)\right).
\end{equation}
Therefore, the static contribution's asymptotics is completely ruled by a damping exponential and thus complies with the vanishing of the Serber potential far from the source. 

\bibliographystyle{unsrtnat}       
\bibliography{libraryneu.bib}

\end{document}